\documentclass[a4paper,twoside]{article}

\usepackage{epsfig}
\usepackage{subfigure}
\usepackage{calc}
\usepackage{amssymb}
\usepackage{amstext}
\usepackage{amsmath}
\usepackage{amsthm}
\usepackage{multicol}
\usepackage{pslatex}
\usepackage{apalike}
\usepackage{psfrag}
\usepackage{flushend}
\usepackage{SCITEPRESS}     
\usepackage[colorlinks=true,hyperfootnotes=true,breaklinks=true]{hyperref}

\begin{document}

\title{Searching for effective and efficient way of~knowledge transfer within an organization}


\author{\authorname{Agnieszka Kowalska-Stycze\'n\sup{1}, Krzysztof Malarz\sup{2} and Kamil Paradowski\sup{2}}
\affiliation{\sup{1}Silesian University of Technology, Faculty of Organization and Management\\ ul. Roosevelta 26/28, PL-41800 Zabrze, Poland}
\affiliation{\sup{2}AGH University of Science and Technology, Faculty of Physics and Applied Computer Science\\ al. Mickiewicza 30, PL-30059 Krakow, Poland}
\email{\href{mailto:Agnieszka.Kowalska-Styczen@polsl.pl}{Agnieszka.Kowalska-Styczen@polsl.pl}, \href{mailto:Krzysztof.Malarz@fis.agh.edu.pl}{Krzysztof.Malarz@fis.agh.edu.pl}}
}


\abstract{In this paper three models of knowledge transfer in organization are considered.
In the first model (A) the transfer of chunks of knowledge among agents is possible only when the sender has exactly one more chunks of knowledge than recipient. This is not dissimilar with bounded confidence model of opinion dynamics. 
In the second model (B) the knowledge transfer take place when sender is ``smarter'' than recipient.
Finally, in the third scenario (model C) we allow for knowledge transfer also when sender posses the same or greater number of chunks of knowledge as recipient.
The simulation bases on cellular automata technique. 
The organization members occupy nodes of square lattice and they interact only with their nearest neighbors.
With computer simulations we show, that the efficiency and the effectiveness of knowledge transfer \emph{i}) for model C is better than for model B \emph{ii}) and it is worse for model A than for model B.}

\onecolumn \maketitle \normalsize \vfill

\section{\label{sec:introduction}\uppercase{Introduction}}

Numerous studies show a relationship between social interaction and knowledge transfer effectiveness and organization performance.
For example, Chen and Huang~\cite{Chen-2007} studied the impact of organizational climate and organizational structure for knowledge management from the perspective of social interaction.
The authors of Ref.~\cite{Chen-2007} emphasize that the competitiveness of companies depends not only on knowledge creation, but mainly from the knowledge diffusion and application of knowledge in the organization.
They also point out that, when the organizational structure is less formal, social interaction is more favorable for the transfer and improves the process of knowledge management.
It was observed also that friendship networks significantly support the transfer of knowledge, because in such situation
\emph{i}) the enhanced cooperation,
\emph{ii}) eased competition
\emph{iii}) and better exchange of information
take place~\cite{Ingram-2000}. 

The influence of the structure of a network of informal contacts on knowledge transfer was also studied by the Reagans and McEvily~\cite{Reagans-2003}.
The study was conducted at R~\&~D sector, which has a very horizontal organizational structure, without a formal hierarchy and which employs mostly scientists and engineers.
The authors of Ref.~\cite{Reagans-2003} analyzed, among others, such factors as: \emph{i}) the ease of knowledge transfer from the source to the recipient, \emph{ii}) codifiability (ease of availability of knowledge for others), \emph{iii}) the strength of ties (the intensity of the connections in terms of the emotional closeness, \emph{iv}) and the frequency of communication).
Their findings indicate that \emph{i}) both social cohesion and long range of interactions facilitate the transfer of knowledge and \emph{ii}) strong ties should be used for the transfer of tacit knowledge.

All these works show the importance of social interaction, especially informal contacts in the process of knowledge transfer.
Moreover, Reagan and McEvily \cite{Reagans-2003} demonstrated, that both strong ties and a dense network of contacts facilitate knowledge transfer.
Such an approach is an indication to the simulation research and bottom-up approach, in which relationships at the local level generate the phenomenon on a global level (i.e. at the level of the whole organization).
Therefore we use the technique of cellular automata (CA) to model the knowledge transfer basing on close relationships between members of the organization.
In our research we focus on the transfer of knowledge within the organization, because, as postulated by Watson and Hewett~\cite{Watson-2006}, acquiring the knowledge contained within the organization is the key \emph{i}) to building a competitive advantage and \emph{ii}) to improve business performance.
New knowledge is created when people communicate and share their knowledge, assimilate it and apply it~\cite{Wah-1999}.
As Davenport and Prusak add, knowledge exists in humans and is an integral part of human complexity and unpredictability \cite{Davenport-2000}.

As pointed out by P\'erez-Nordtvedt et~al.~\cite{Perez-2008}, a complete picture of knowledge transfer yield two dimensions of knowledge transfer: \emph{the effectiveness} and \emph{the efficiency}.
Therefore, we study different ways of transfer knowledge and we are looking for the most efficient and effective way of transferring knowledge in the organization.
It should be noted that the efficiency is understood as the level of adoption of knowledge by the entity receiving~\cite{Jensen-2007}, and the effectiveness is defined as the speed with which the recipient acquires new knowledge~\cite{Perez-2008}.

The starting point for our research is recently proposed a CA model~\cite{Kowalska-2017}, in which---basing on empirical findings~\cite{Reagans-2003}---a small distance of common knowledge between the source and recipient was assumed.
Such approach is consistent with Deffuant et al. model~\cite{Deffuant-2000} of opinion dynamics where opinion exchange among agents is possible only when sender and recipient have similar opinions~\cite{Hegselmann-2002,Malarz2006b,Kulakowski-2009,Gronek2011,Kulakowski2014}.

The authors of Ref.~\cite{Kowalska-2017} assumed, that agents (members of the organization) can acquire knowledge only of the neighbors, who have one more piece of knowledge from them.
This causes that at some point the agents cannot acquire more knowledge, because everyone in their neighborhood are ``too smart'' for them. 
Therefore, the transfer of the knowledge is not fully effective, and this suggests the search for more efficient and effective way of transfer knowledge within the organization and for looking for improved models of this phenomenon.

\section{\uppercase{Model}}

\begin{figure*}[!htbp]
\centering
\psfrag{n(k)}{$n(k) [\%]$}
\psfrag{t}{$t$}
\psfrag{k}{$k=$}
\psfrag{1-1: L=20, K=4, p=0.200}[][c]{(a) model A: $p=0.2$}
\psfrag{1-1: L=20, K=4, p=0.500}[][c]{(b) model A: $p=0.5$}
\psfrag{1-K: L=20, K=4, p=0.2}[][c]{(c) model B: $p=0.2$}
\psfrag{1-K: L=20, K=4, p=0.5}[][c]{(d) model B: $p=0.5$}
\psfrag{0-K: L=20, K=4, p=0.2}[][c]{(e) model C: $p=0.2$}
\psfrag{0-K: L=20, K=4, p=0.5}[][c]{(f) model C: $p=0.5$}
\includegraphics[width=0.30\textwidth]{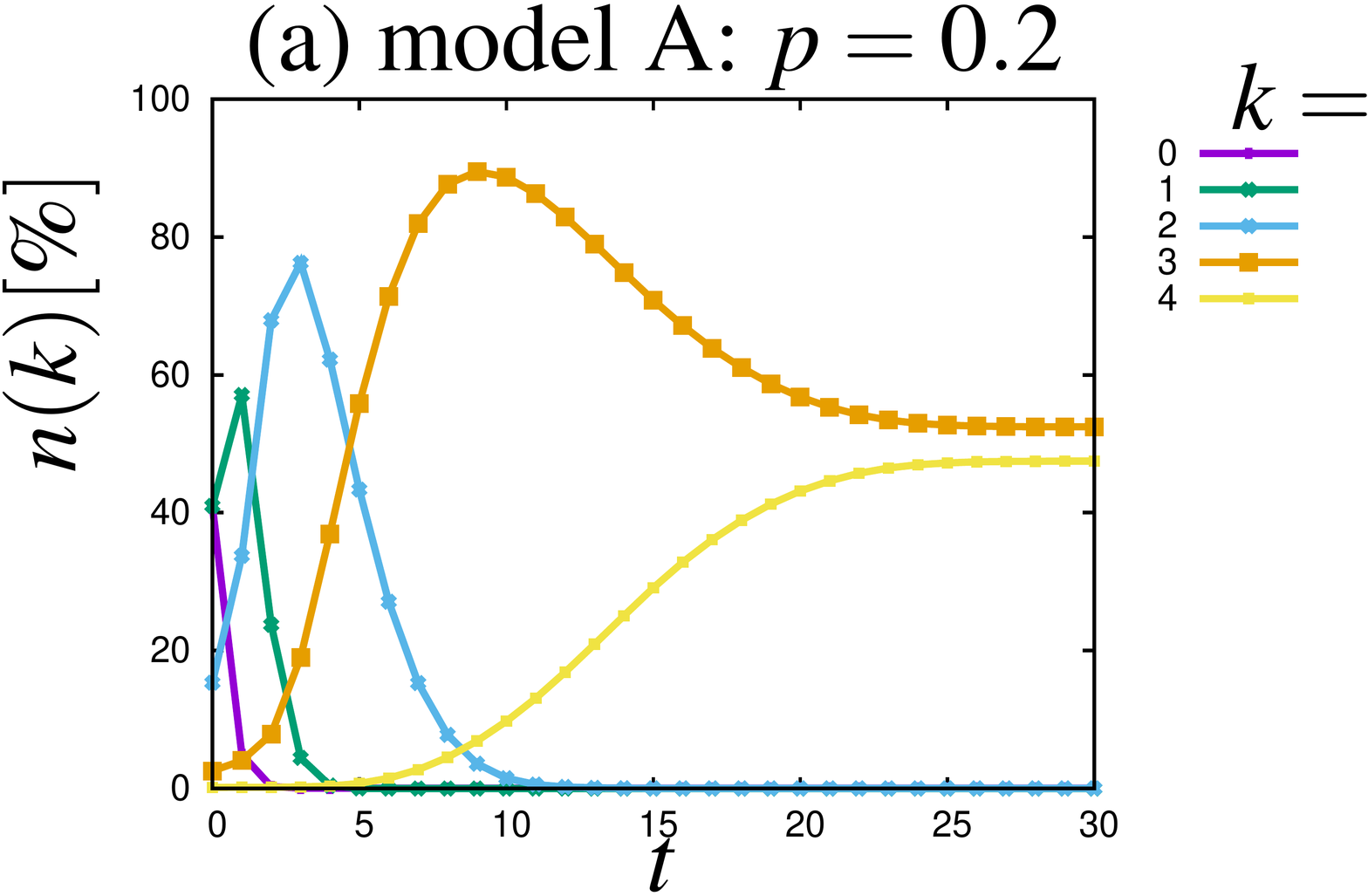}
\includegraphics[width=0.30\textwidth]{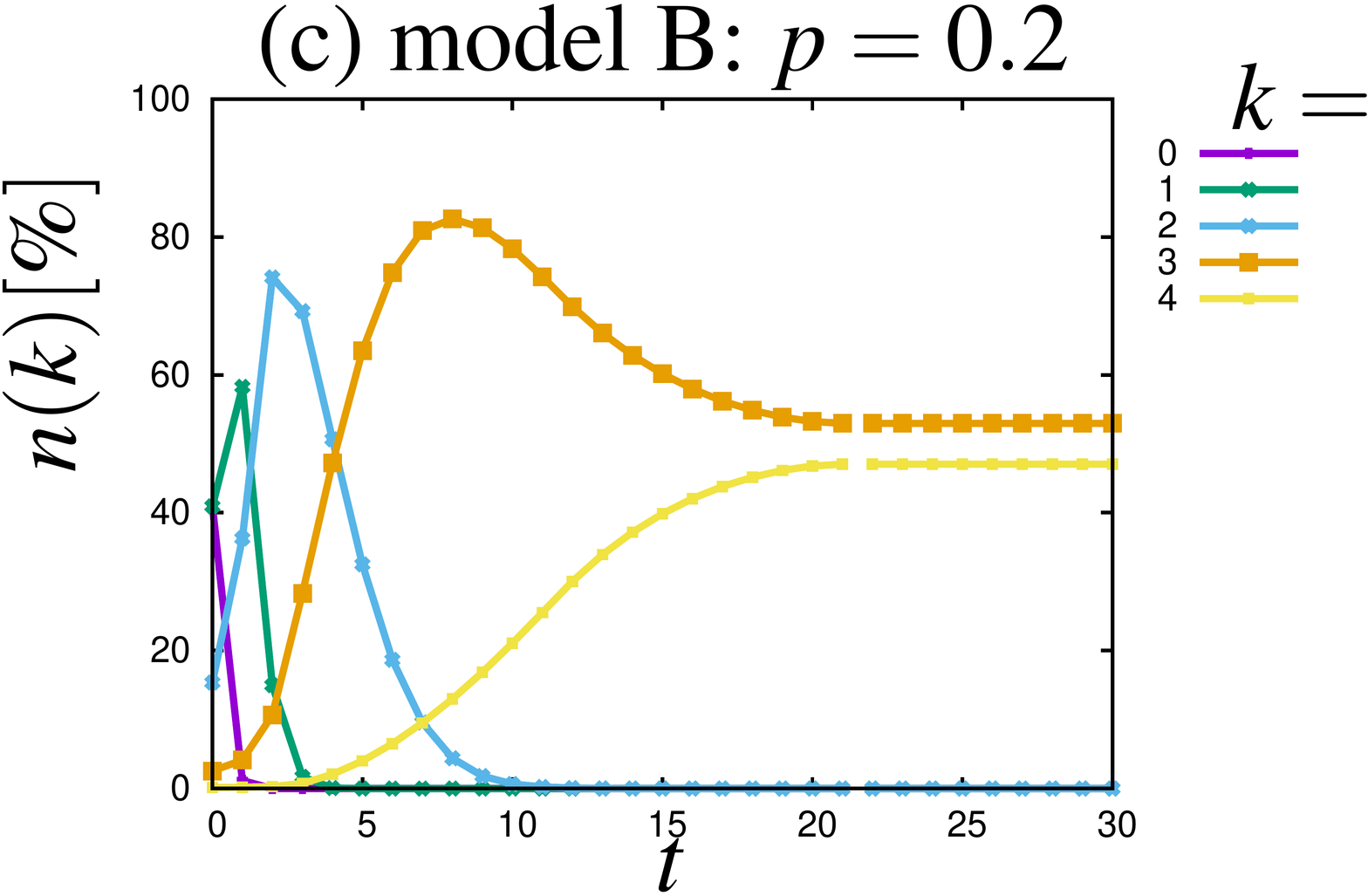}
\includegraphics[width=0.30\textwidth]{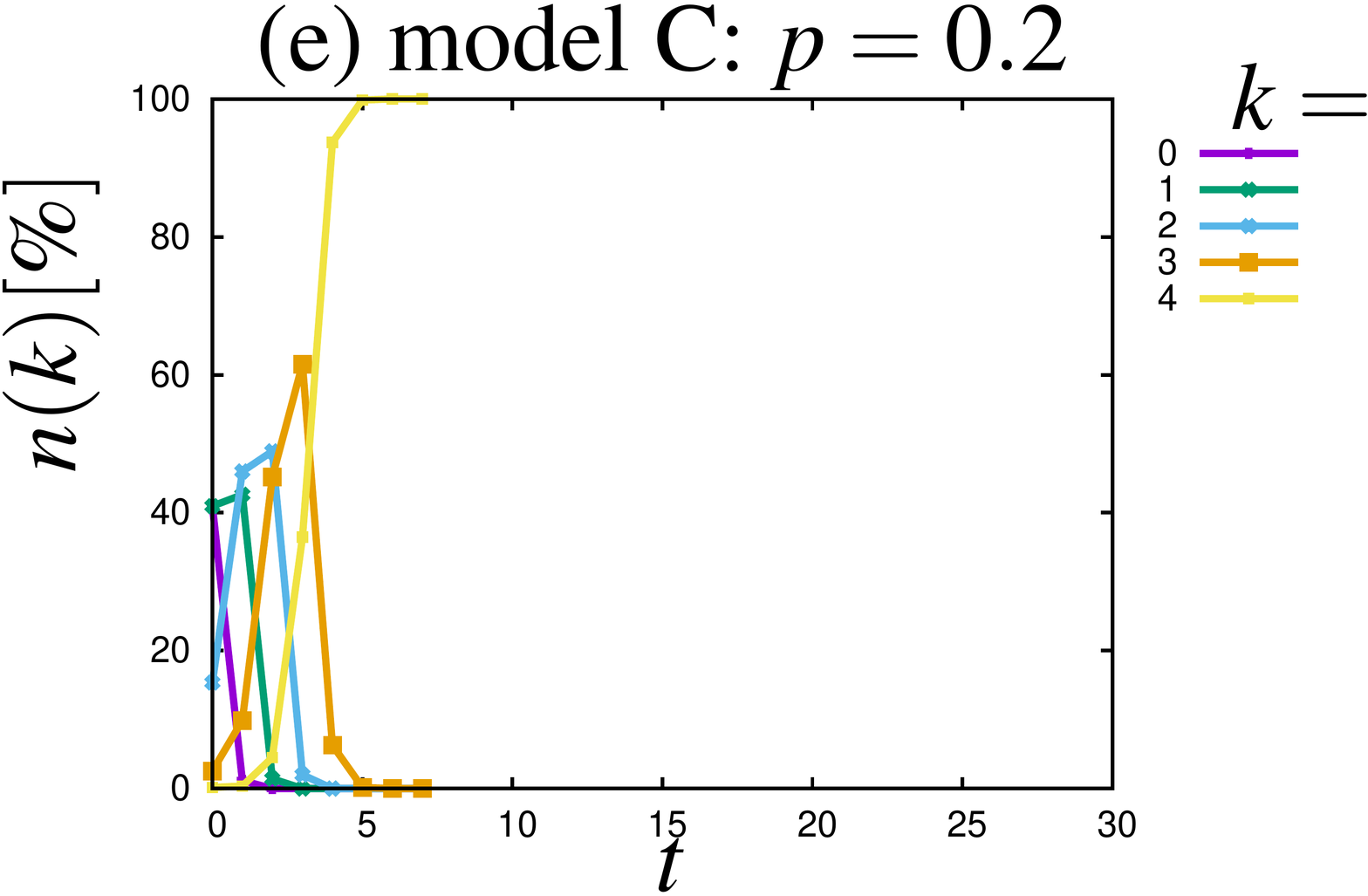}\\[2mm]
\includegraphics[width=0.30\textwidth]{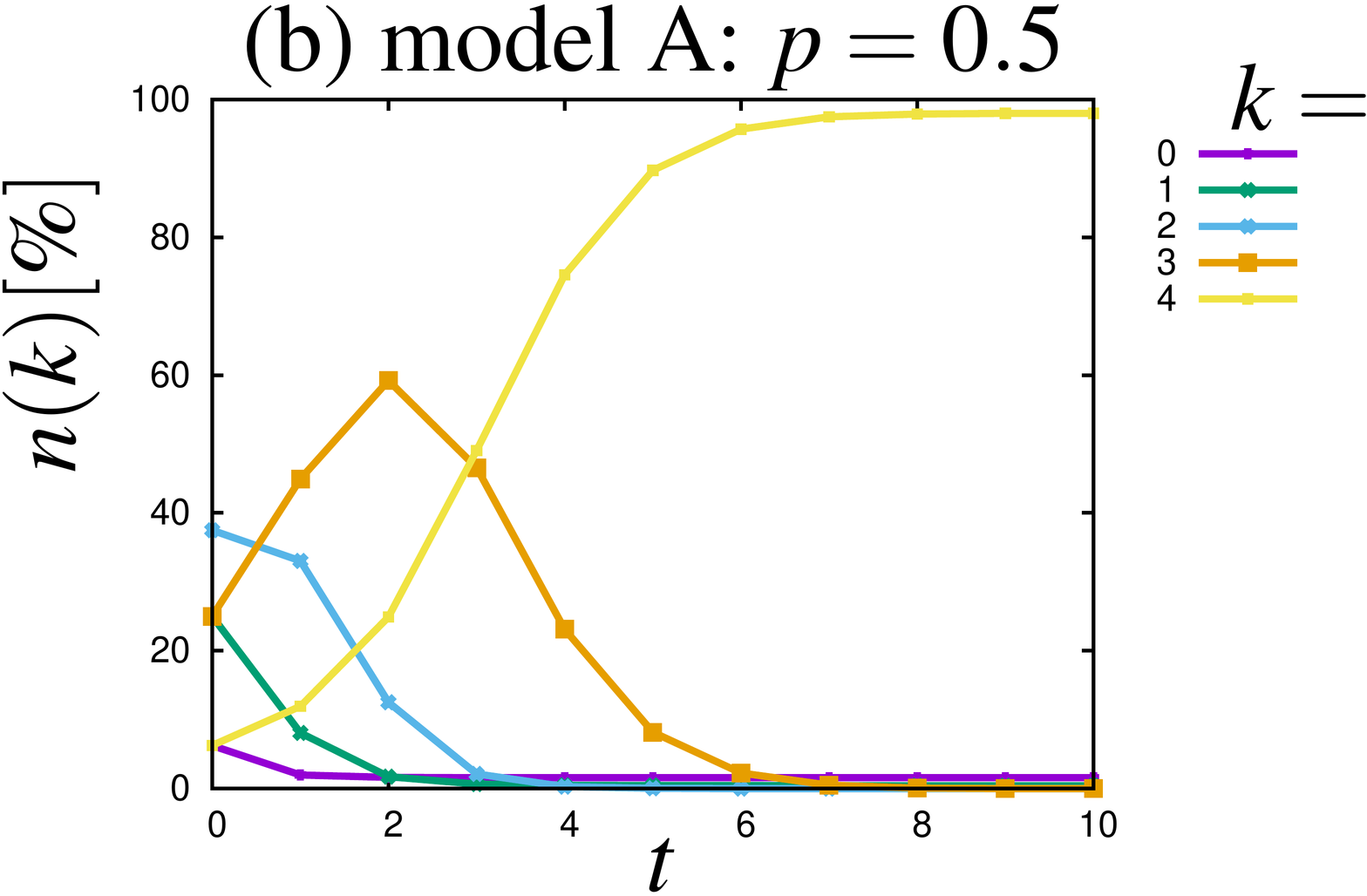}
\includegraphics[width=0.30\textwidth]{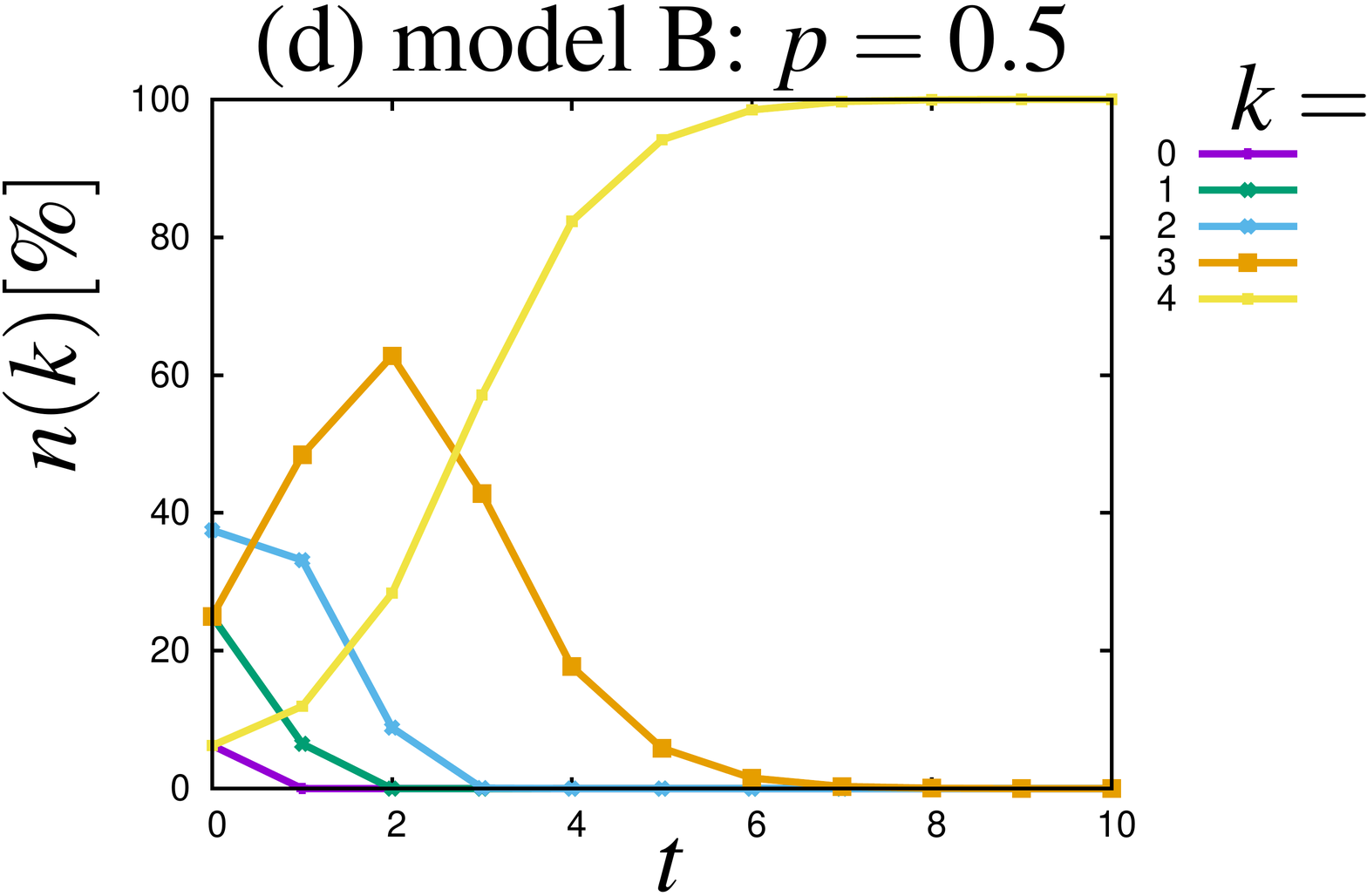}
\includegraphics[width=0.30\textwidth]{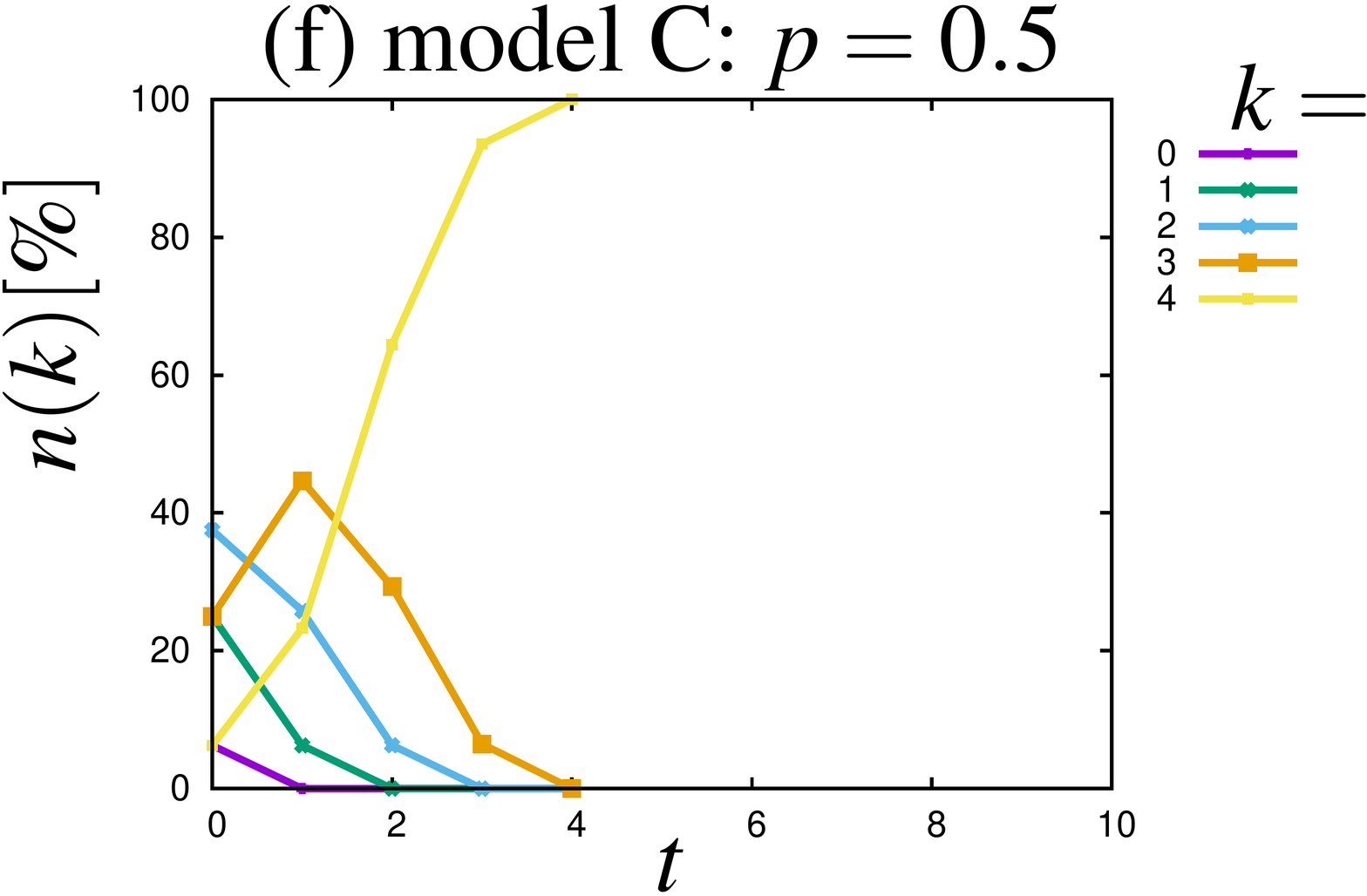}
\caption{\label{F:nkomp_vs_t} The time evolution of the fraction $n(k)$ of agents having $k$ chunks of knowledge for $L=20$, $K=4$ and various initial concentration $p$ of chunks of knowledge among agents.
The results are averaged over $M=10^4$ independent simulations.
(a)-(b) model A, (c)-(d) model B, (e)-(f) model C.}
\label{fig_sim}
\end{figure*}

In Ref.~\cite{Kowalska-2017} a CA to simulate the knowledge transfer within an organization was defined.
To describe a CA one should define 
\emph{i}) a $d$-dimensional network,
\emph{ii}) the set of states of network nodes
\emph{iii}) and the rule defining the system dynamics.
The above mentioned rule defines state of node in discrete time $t+1$ basing on this node and its neighborhood state in time $t$.
Here we assume a two-dimensional square lattice with von Neumann neighborhood, i.e. node (agent) at discrete coordinate $(x,y)\in\mathbb{N}\times\mathbb{N}$ and $1\le x,y \le L$ has only four nearest neighbors at $(x',y')\in\{(x-1,y), (x+1,y), (x,y-1), (x,y+1)\}$.
The set of states describes chunks of knowledge possessed by the agent.
Namely, the vector variable 
\[
\mathbf{C}(x,y,t)=[c_1(x,y,t),c_2(x,y,t),\cdots,c_K(x,y,t)]
\]
indicates if agent at time $t$ and at the coordination $(x,y)$ posses $k$-th chunk of knowledge [$c_k(x,y,t)=1$] or not [$c_k(x,y,t)=0$] among $K$ chunks of knowledge desired by organization for their members.
Initially (at $t=0$), each agent has each chunk of knowledge with probability $p$.
Every time step $t$ agent at $(x,y)$ may inherit {\em single} chunk of knowledge from {\em one} of its randomly selected neighbor.

\subsection{\label{modelA}Model A}

For model proposed in Ref.~\cite{Kowalska-2017} the transfer of chunks of knowledge among agents is possible only when the sender has \emph{exactly one more} chunks of knowledge than recipient:
\begin{multline}
\label{eq:A}
c_k(x,y,t+1)=1 \iff c_k(x,y,t)=0\\ \text{ and } c_k(x',y',t)=1 \text{ and } \nu(x',y',t)=\nu(x,y,t)+1,
\end{multline}
where $\nu(x,y,t)$ and $\nu(x',y',t)$ are numbers of chunks of knowledge at time $t$ for receiver and sender of knowledge, respectively.

\subsection{\label{modelB}Model B}

As we mentioned in the Introduction, here we would like to omit the restriction of knowledge transfer only from senders having exactly one more chunk of knowledge than recipient. 
Namely, we allow for sending chunk of knowledge from neighbors at $(x',y')$ to recipient at $(x,y)$ if sender has {\em more} chunks of knowledge than recipient:
\begin{multline}
\label{eq:B}
c_k(x,y,t+1)=1 \iff c_k(x,y,t)=0\\ \text{ and } c_k(x',y',t)=1 \text{ and } \nu(x',y',t)>\nu(x,y,t).
\end{multline}

\subsection{\label{modelC}Model C}

Finally, we modify the rule described in Sec.~\ref{modelB} in order to allowing for transfer of chunks of knowledge among agents with \emph{the same or grater} numbers of chunks of knowledge:
\begin{multline}
\label{eq:C}
c_k(x,y,t+1)=1 \iff c_k(x,y,t)=0\\ \text{ and } c_k(x',y',t)=1 \text{ and } \nu(x',y',t)\ge\nu(x,y,t).
\end{multline}

In further part of this paper we will refer to these three models as model A [Eq.~\eqref{eq:A}], model B [Eq.~\eqref{eq:B}] and model C [Eq.~\eqref{eq:C}], respectively.

\section{\uppercase{The design of experiments}}

Experiments that have been conducted were designed to investigate which of the three models gives the most effective and efficient transfer of knowledge. 

Similarly to the work~\cite{Kowalska-2017}, the following independent variables were chosen in the designed experiments:
\emph{i}) maximal number of chunks of knowledge $K$
\emph{ii}) and initial concentration of chunks of knowledge $p$.

Our goal is to seek the most efficient and the most effective way to transfer knowledge.
In view of the above, we define the corresponding variables describing these two parameters of knowledge transfer.
Firstly, as the dependent variable describing an effectiveness of knowledge transfer the following parameters were adopted:
\begin{itemize}
\item the fraction $n(k)$ of agents having $k$ chunks of knowledge,
\item the fraction $n(K)$ of agents having all chunks of knowledge,
\item the fraction $f(k)$ of agents having $k$-th chunk of knowledge $c_k$,
\item the coverage $\langle f\rangle$ of any chunks of knowledge $c_k$ in agents’ knowledge, i.e. the fraction of knowledge chunks held by typical member of the organization.
\end{itemize}
Secondly, as the dependent variable describing an efficiency of knowledge transfer, the time $\tau$ necessary for reaching a stationary state was adopted.

\section{\uppercase{Results}}

In order to compare efficiency of the knowledge transfer among newly proposed (Sec.~\ref{modelB} and~\ref{modelC}) and previously discussed (Sec.~\ref{modelA}) models we have reproduced some results presented in Ref.~\cite{Kowalska-2017} but for ten times better statistic, i.e. for $M=10^4$, where $M$ is the number of independent simulations used for averaging procedure.
This number of independent simulations is also applied for simulations based on models B and C.

\subsection{\label{sec:nk}Time evolution of the fraction of agents having $k$ chunks of knowledge}

In Fig.~\ref{F:nkomp_vs_t} the time evolution of the fraction $n(k)$ of agents having $k$ chunks of knowledge for $L=20$, $K=4$ and $p=0.2$, 0.5 and for various models are presented.
The fraction $n(k)$ is formally defined as
\begin{equation}
n(k)\equiv\dfrac{\sum_{r=1}^M N_r(k)}{ML^2},
\end{equation}
where $N_r(k)$ stands for the number of agents having exactly $k$ chunks of knowledge in $r$-th simulation, $M$ is the number of independent simulations and $L$ is linear size of the system.
For low initial concentration of chunks of knowledge among agents ($p=0.2$) the differences between results of model A and model B seems to be negligible, except that time $\tau$ of reaching the stationary state is a little bit shorter for model B.
The same apply for $p=0.5$.
Here however, additionally we observe that fraction of agents possessing less than $K$ chunks of knowledge vanishes for long enough simulation times.
Simultaneously, the fraction of agents with all chunks of knowledge saturates at $n(K)=1$ while $n(K)<1$ for simulations based on model A.
These effects are enhanced when model C is applied.
Moreover, for model C the fraction of agents with all chunks of knowledge increases to $n(K)=1$ even for low initial level ($p=0.2$) of chunks of knowledge among agents.

The effectiveness of the knowledge transfer measured by $n(k)$ depends on $p$ and on the method of knowledge transfer (models A, B, C) as presented in Fig.~\ref{F:nkomp_vs_t}.
The greater initial concentration $p$ of chunks of knowledge the higher the percentage $n(k)$ of agents having $k$ chunks of knowledge, as can be seen by comparing upper [Fig.~\ref{F:nkomp_vs_t}(a, c, e), $p=0.2$] and lower [Fig.~\ref{F:nkomp_vs_t}(b, d, f), $p=0.5$] panels of Fig.~\ref{F:nkomp_vs_t}.
Furthermore, comparing the results for model A, B and C one can see that the knowledge transfer is most effective in the case of the model C while it is quite similar in the case of models A and B.

Now let us look at the efficiency of knowledge transfer.
In Fig.~\ref{F:nkomp_vs_t} the efficiency can be understood as the time (number of simulation steps) required to obtain the maximum fraction $n(k)$ of agents having $k$ chunks of knowledge.
In the case of the model C, the time of the transfer of knowledge is much shorter than for models A and B.

\subsection{\label{sec:nK}Time evolution of the fraction of agents having all chunks of knowledge}

\begin{figure}[!htbp]
\centering
\psfrag{1-1: L=20, K=4}[][c]{(a) model A}
\psfrag{1-K: L=20, K=4}[][c]{(b) model B}
\psfrag{0-K: L=20, K=4}[][c]{(c) model C}
\psfrag{p=}{\small {$p=$}}
\psfrag{t}{$t$}
\psfrag{n(K)}[][c]{$n(K)$ [\%]}
\includegraphics[width=.45\textwidth]{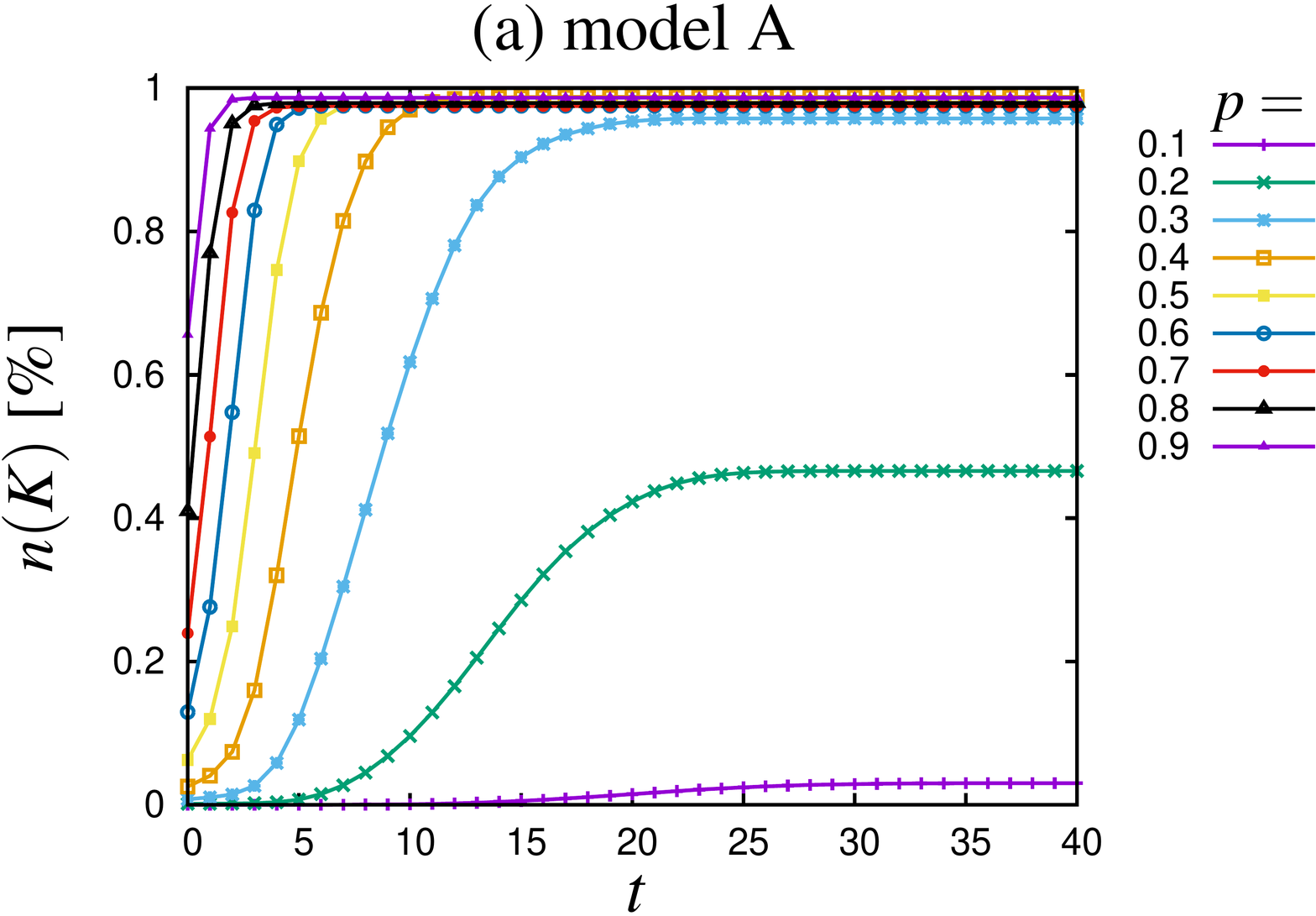}
\includegraphics[width=.45\textwidth]{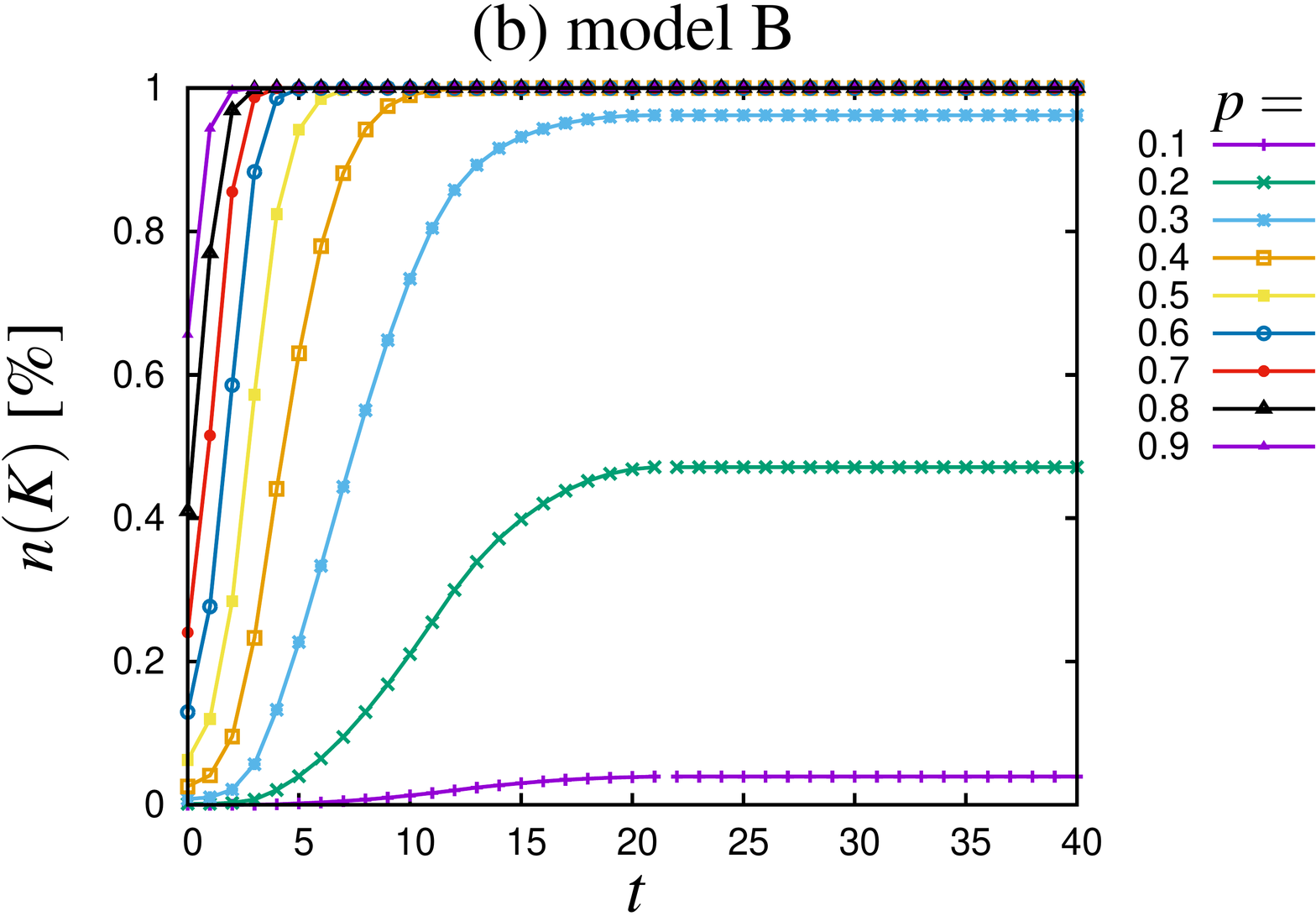}
\includegraphics[width=.45\textwidth]{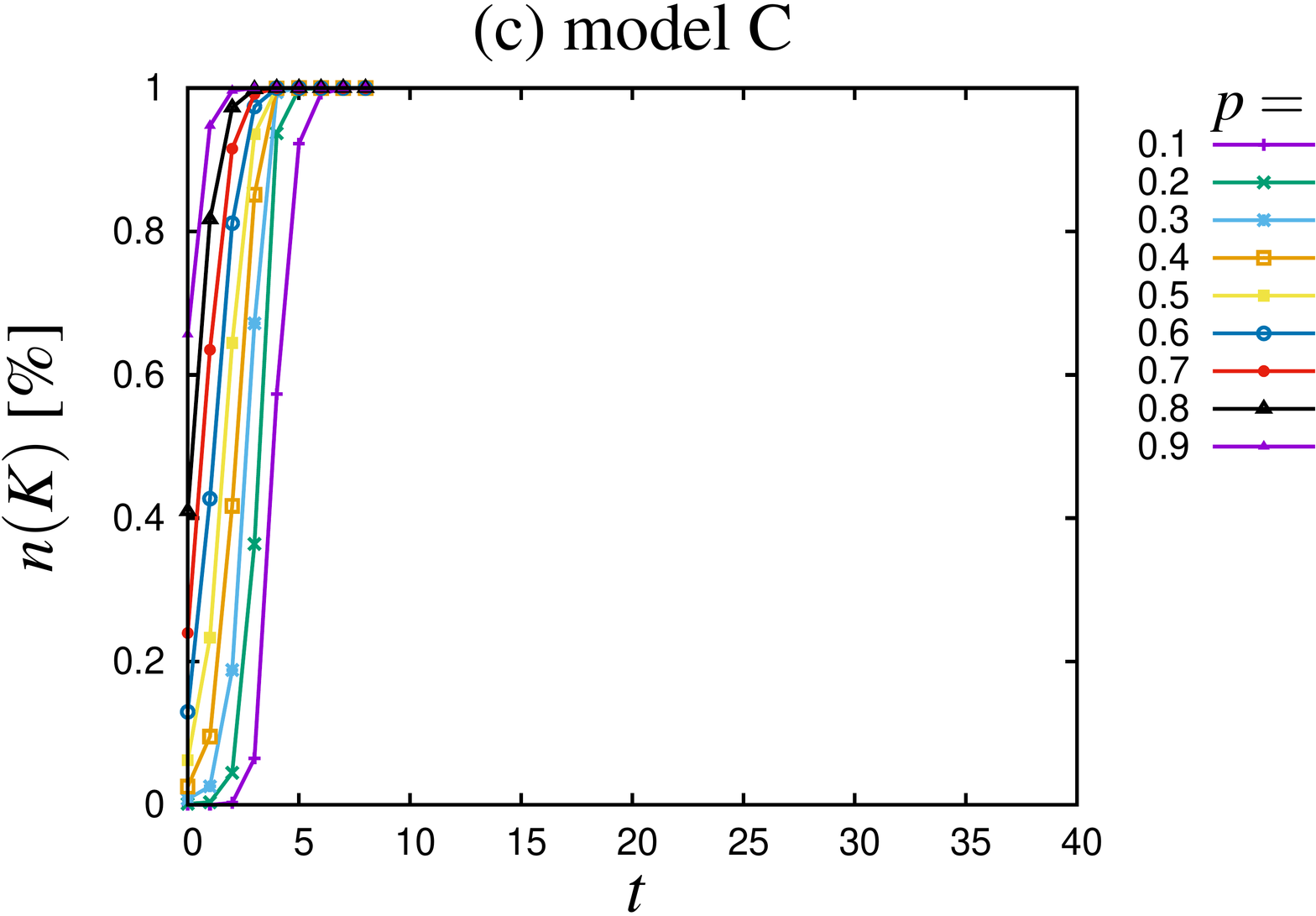}
\caption{\label{F:omni_vs_t}The time evolution of the fraction $n(K)$ of agents having total knowledge---i.e. possessing all $K$ chunks of knowledge $(c_1,c_2,\cdots,c_K)=(1,1,\cdots,1$) for $L=20$ and $K=4$.
The values of $n(K)$ are averaged over $M=10^4$ independent simulations.}
\end{figure}

In Fig.~\ref{F:omni_vs_t} the time evolution of the fraction $n(K)$ of agents having total knowledge---i.e. possessing all $K$ chunks of knowledge $(c_1,c_2,\cdots,c_K)=(1,1,\cdots,1$) is presented.
In Fig.~\ref{F:omni_vs_t}(a) the data for model A are reproduced.
The data obtained for model B [Fig.~\ref{F:omni_vs_t}(b)] exhibit again shorter times of reaching the saturation of curves $n(K)$ with simulation steps $t$ vs. the same data for model A.
Moreover, for model B and for initial concentrations of chunks of knowledge $p>0.3$ the curves $n(K)$ saturates at $n(K)=1$ while in model A $n(K)<1$ independently on $p$.
Finally, for model C, these effects are strongly enhanced and for all checked values of initial concentration of chunks of knowledge $p$ only several simulations steps are required to reach $n(K)=1$.

As it was mentioned earlier, $n(K)$ is also a variable describing the effectiveness of knowledge transfer.
As can be seen in Fig.~\ref{F:omni_vs_t}, the effectiveness of knowledge transfer is greater for model A than B, but definitely the best results gives model C.
In this case, even small values of initial fraction $p$ cause that all the organization members have all $K$ required chunks of knowledge.

\subsection{\label{sec:tau}Time necessary for reaching the stationary state}

\begin{figure*}[!htbp]
\centering
\psfrag{1-1: L=20}[][c]{(a) model A} 
\psfrag{1-K: L=20}[][c]{(b) model B} 
\psfrag{0-K: L=20}[][c]{(c) model C} 
\psfrag{p}{$p$}
\psfrag{tau}{$\tau$}
\psfrag{K=}{{\small $K=$}}
\includegraphics[width=.32\textwidth]{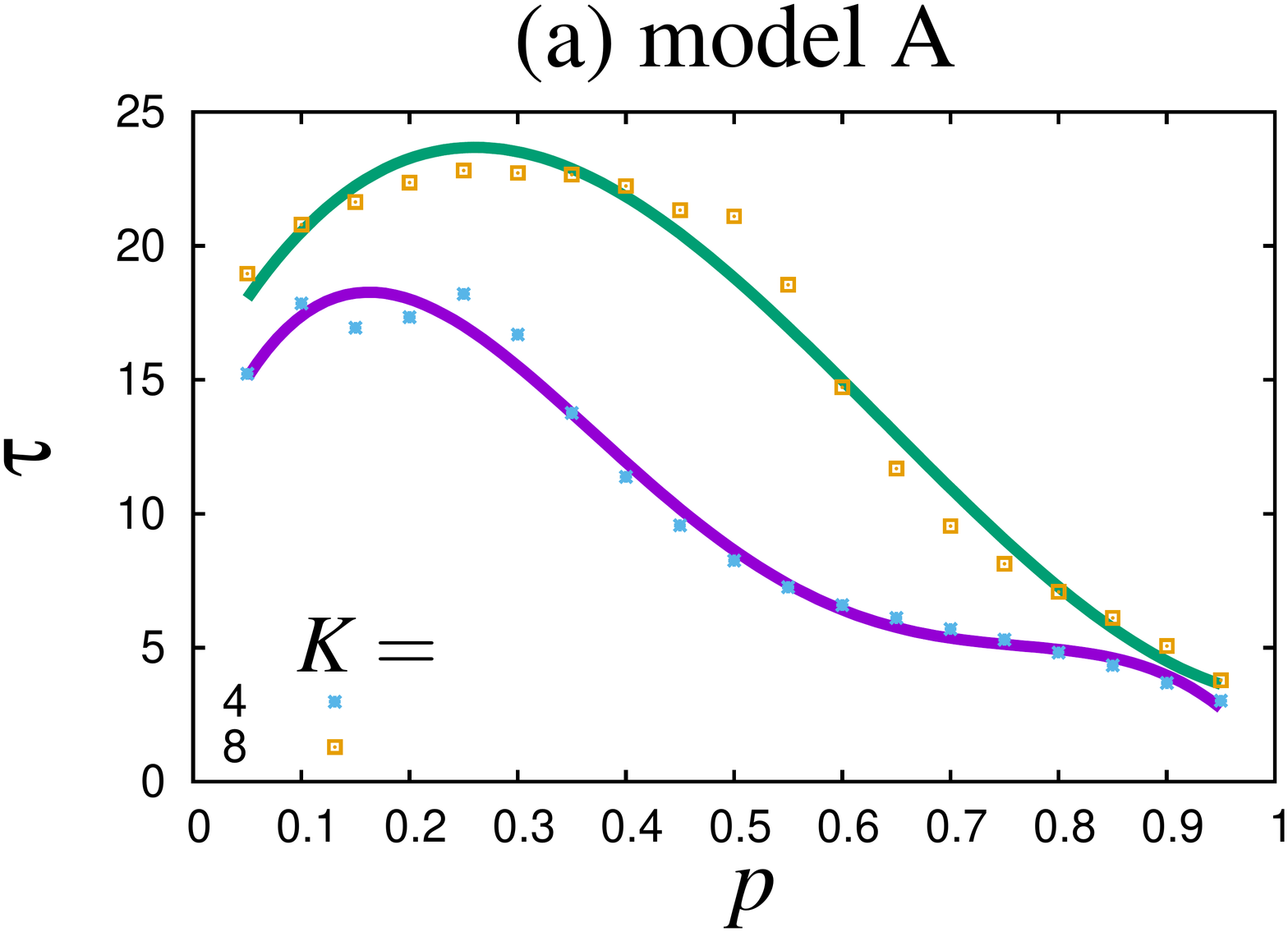}
\includegraphics[width=.32\textwidth]{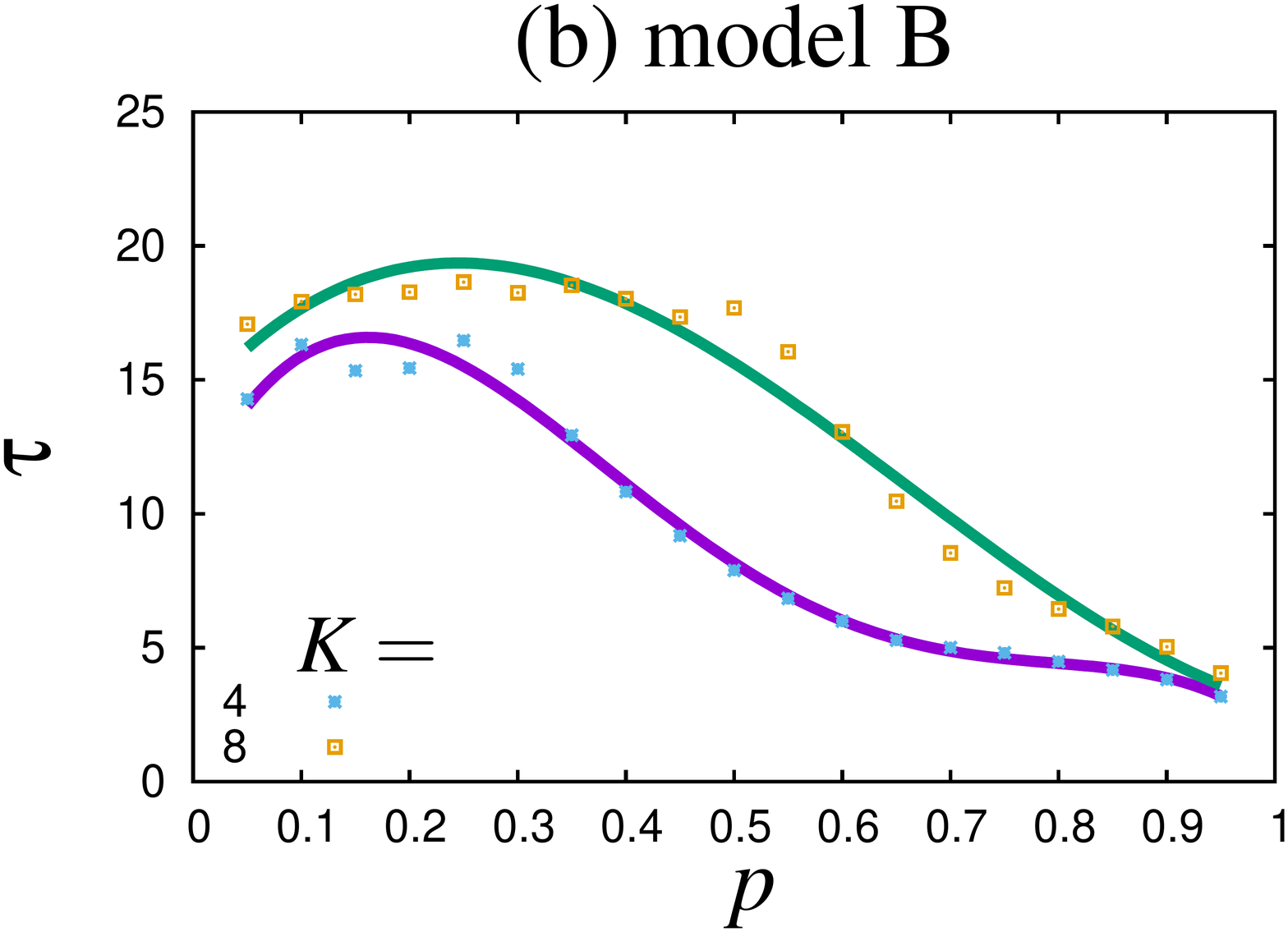}
\includegraphics[width=.32\textwidth]{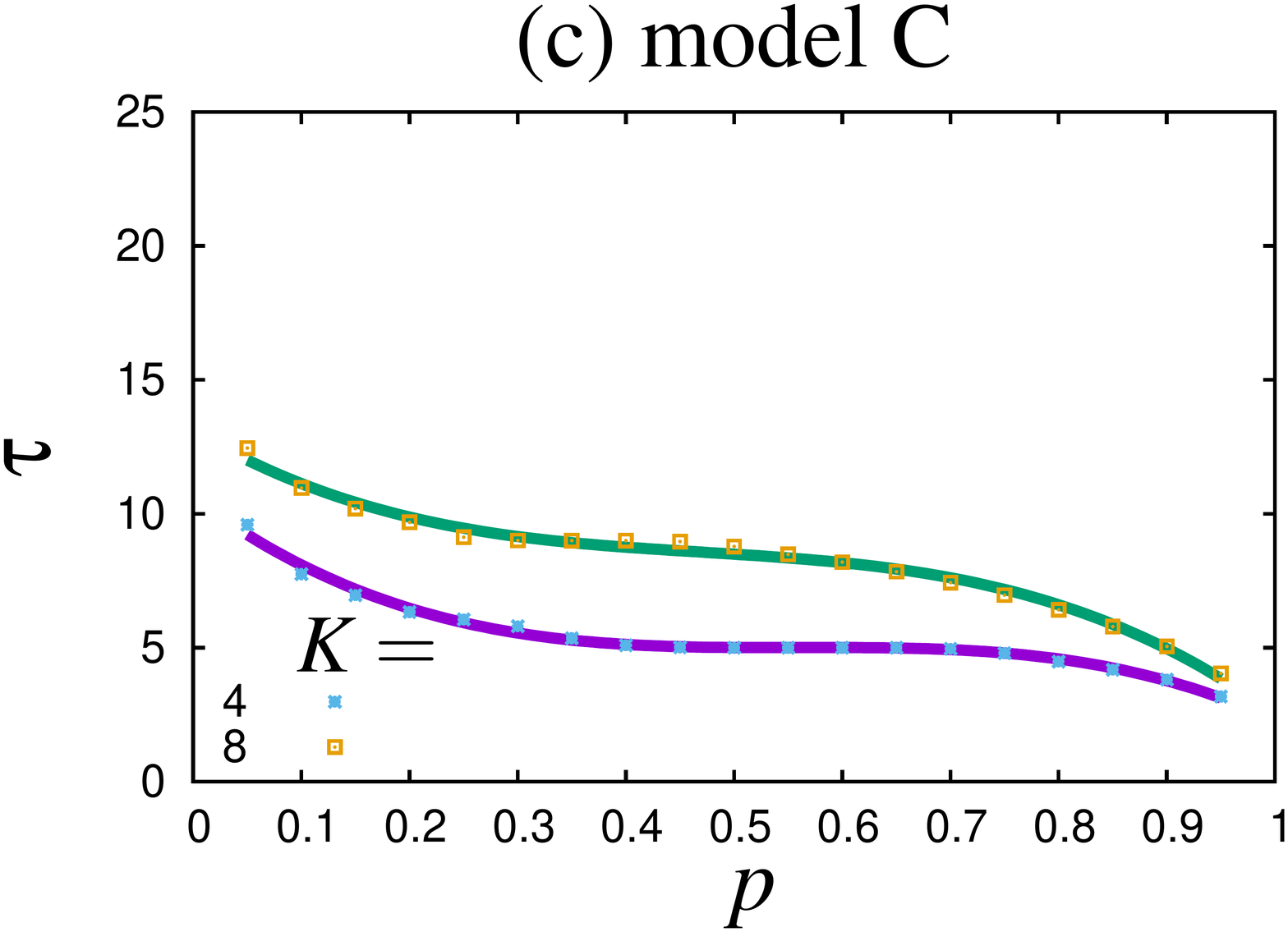}
\caption{\label{F:tau_vs_p}The time $\tau$ necessary for reaching the stationary state as dependent on initial concentration of chunks of knowledge $p$.
The results are averaged over $M=10^4$ independent simulations.
The error bars are smaller than the symbols.
(a) model A, (b) model B, (c) model C.}
\end{figure*}

In Fig.~\ref{F:tau_vs_p} times $\tau$ necessary for reaching the stationary state as dependent on initial concentration of chunks of knowledge $p$ for $L=20$ and $K=4$, 8  and various models are presented.
In Fig.~\ref{F:tau_vs_p}(a) the results presented in Ref.~\cite{Kowalska-2017} are reproduced.
The heuristic low degree polynomial fit has been added for eyes guidance as well.
For model B---when agents may receive chunks of knowledge from ``smarter'' agents---we observe shorter times $\tau$ in respect to results of model A for both checked values of $K$.
However, the characteristic of the fit shapes are roughly the same for models A and B.
This situation changes qualitatively and quantitatively when model C is applied---the obtained values of $\tau$ are twice smaller when comparing with results for model A and the fit shape is different than for models A and B.

The number of simulation steps $\tau$ to obtain the steady state, expresses the efficiency of knowledge transfer.
As can be seen in Fig.~\ref{F:tau_vs_p}, the time of the transfer of knowledge is shorter in the case of model B than A, but definitely it is shortest for the model C.
In the latter case, the time $\tau$ decreases with increasing $p$.
The transfer of knowledge chunks among agents with the same or greater numbers of chunks of knowledge therefore gives the best results.

\subsection{\label{sec:fci}The average coverage of chunks of knowledge in whole organization}

\begin{figure*}[!ht]
\centering
\psfrag{p}{$p$}
\psfrag{K=}{$K=$}
\psfrag{L=20}{(a) model A} 
\psfrag{L=20, top}{(b) model A} 
\psfrag{1-K: L=20}{(c) model B} 
\psfrag{1-K: L=20, top}{(d) model B} 
\psfrag{0-K: L=20}{(e) model C} 
\psfrag{0-K: L=20, top}{(f) model C} 
\psfrag{<fci>}[c]{$\langle f\rangle$ [\%]}
\includegraphics[width=.32\textwidth]{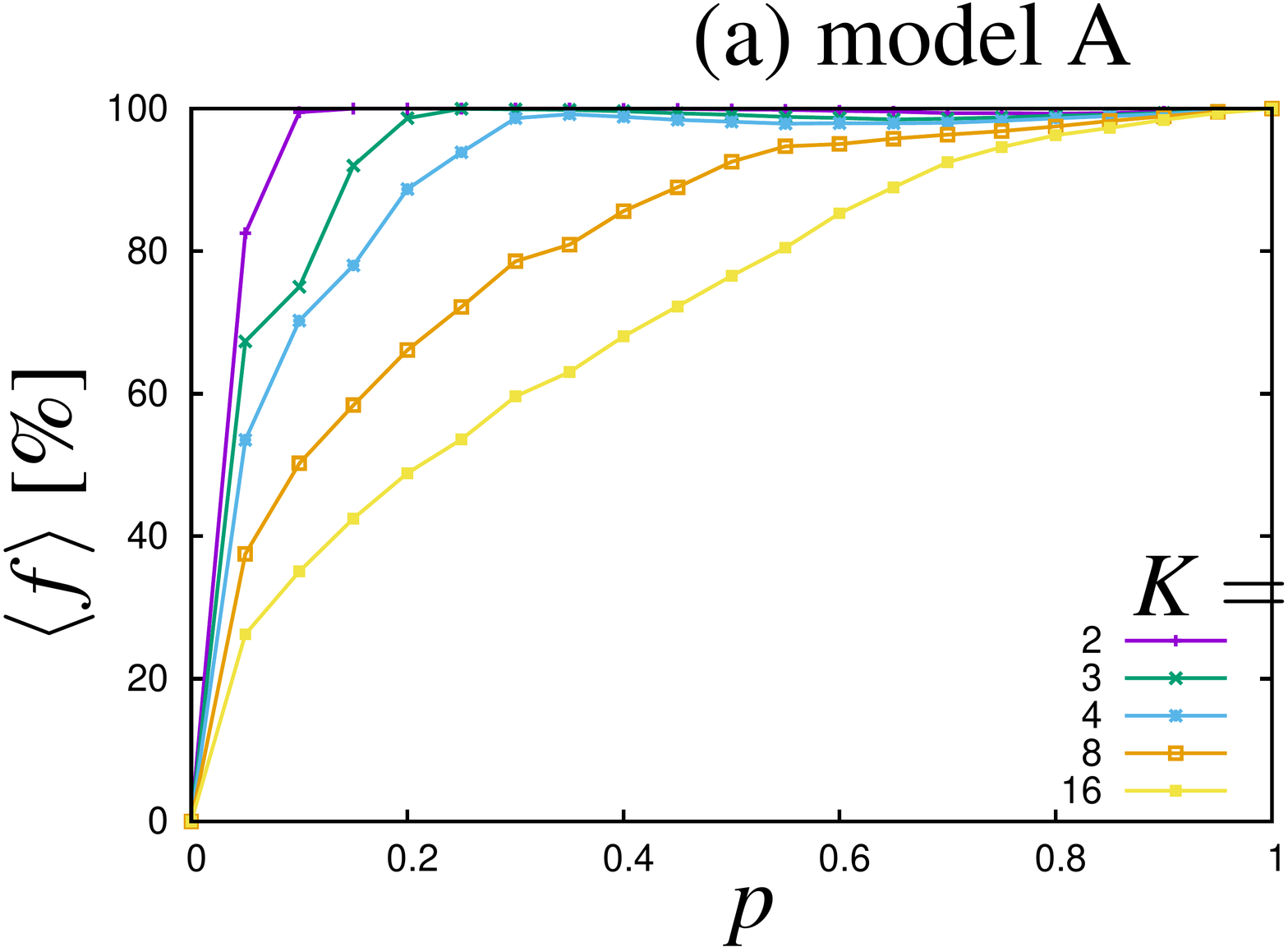}
\includegraphics[width=.32\textwidth]{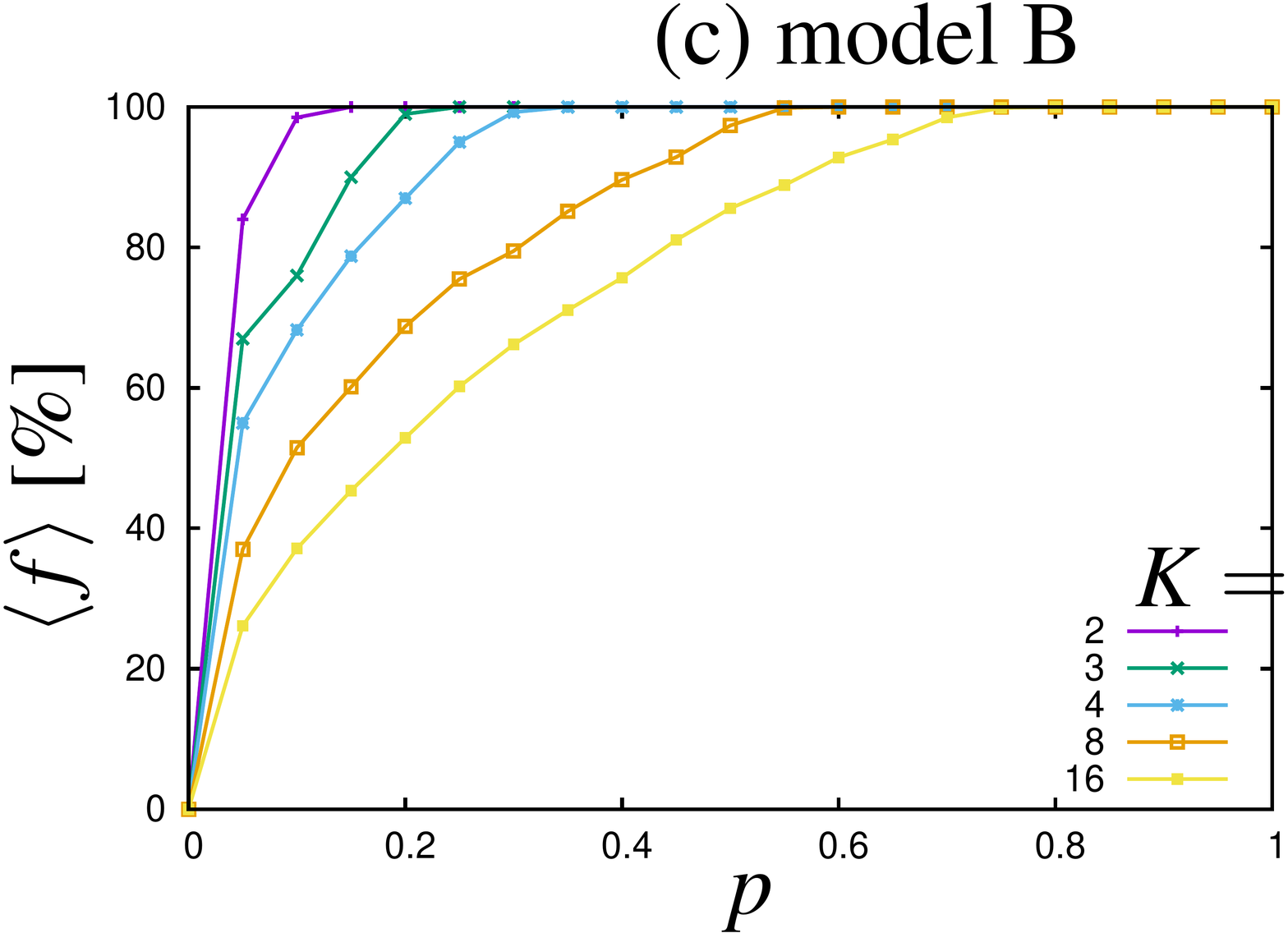}
\includegraphics[width=.32\textwidth]{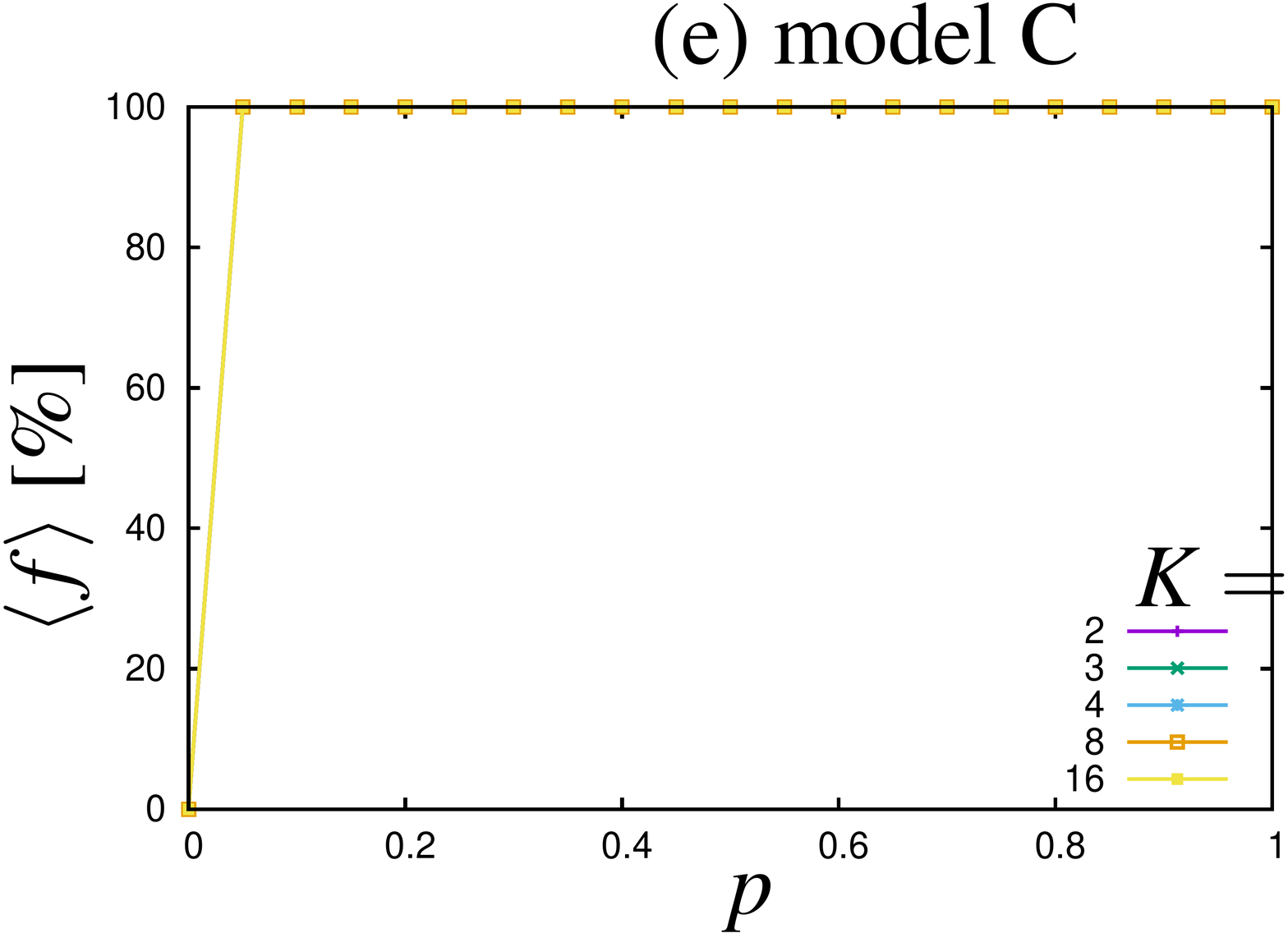}\\[2mm]
\includegraphics[width=.32\textwidth]{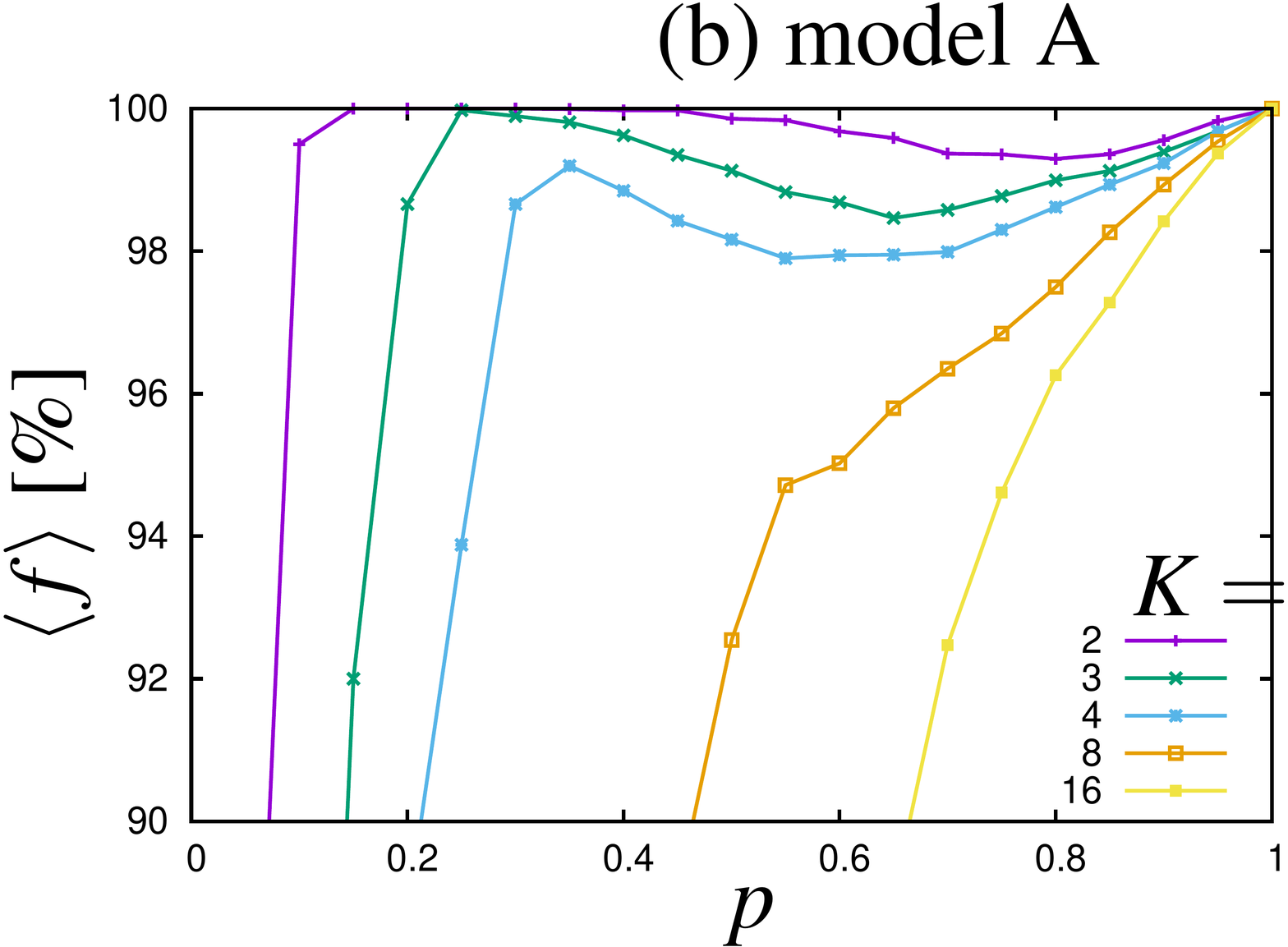}
\includegraphics[width=.32\textwidth]{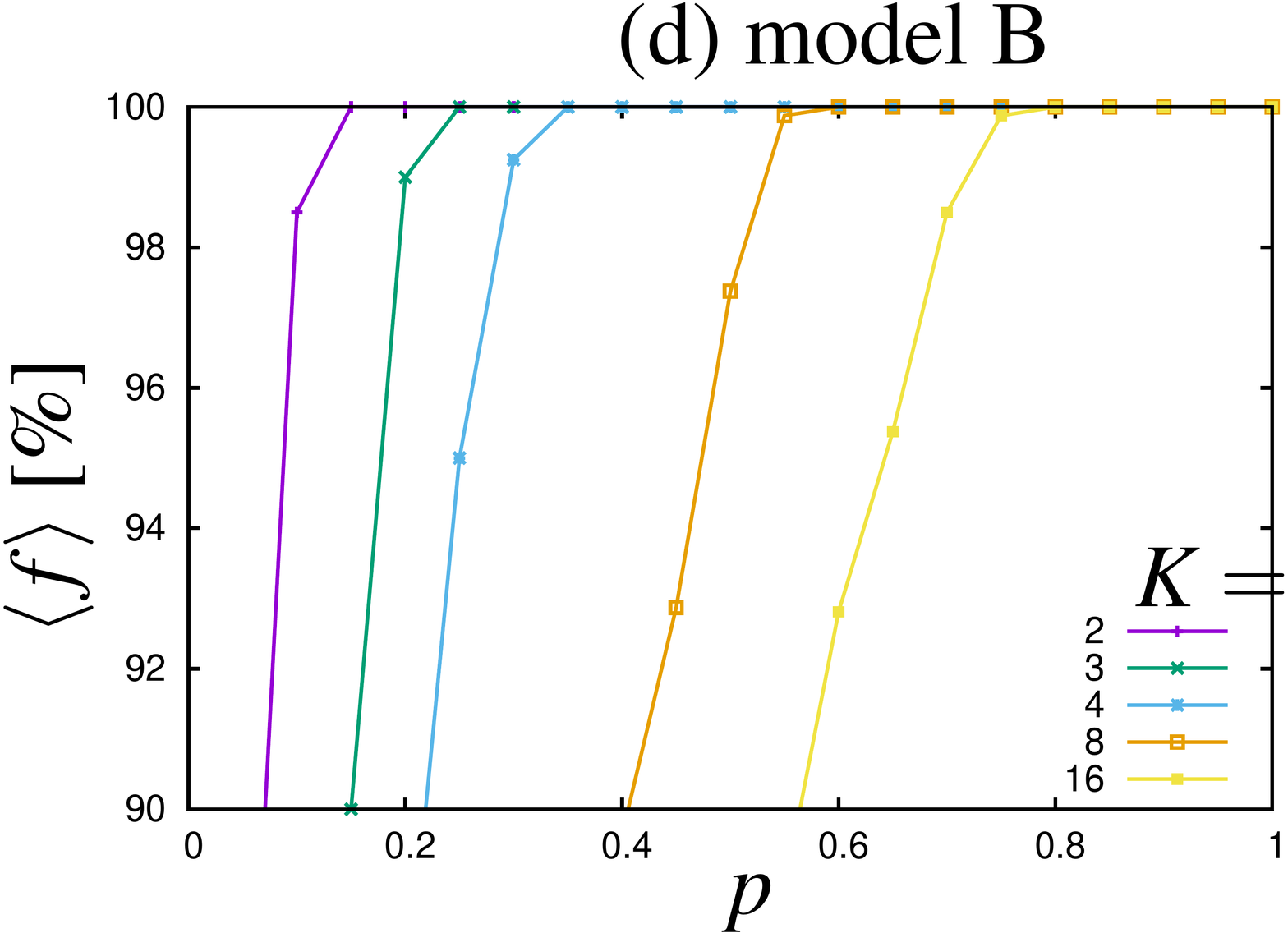}
\includegraphics[width=.32\textwidth]{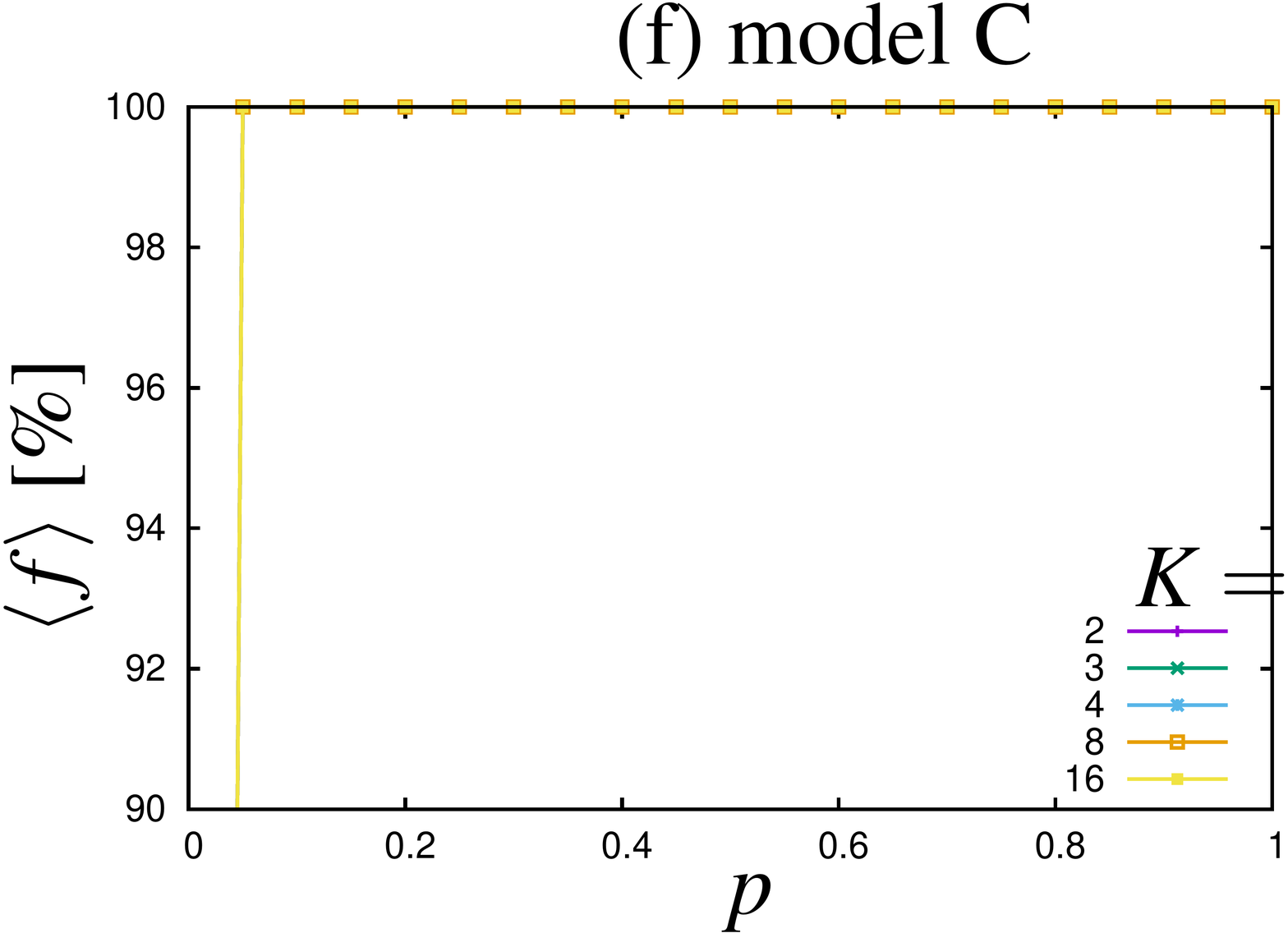}
\caption{\label{F:fci_vs_Kp}The average coverage of chunks of knowledge $\langle f\rangle$ of agents having any chunk of knowledge $c_k$ for $L=20$ and long simulations time $t\to\infty$ and various values of $K$ ($K=2$, 3, 4, 8, 16) as dependent on initial concentration of chunks of knowledge $p$.
The values of $\langle f\rangle$ are averaged over $M=10^4$ independent simulations.}
\end{figure*}

In Fig.~\ref{F:fci_vs_Kp} the average coverage of chunks of knowledge 
\[
\langle f\rangle=K^{-1}\sum_{k=1}^K f(k)
\]
of agents having any chunk of knowledge $c_k$ for $L=20$ and long simulations time $t\to\infty$ and various values of $K$ ($K=2$, 3, 4, 8, 16) as dependent on initial concentration of chunks of knowledge $p$ is presented.
The fraction $f(k)$ is formally defined as
\begin{equation}
f(k)\equiv\dfrac{\sum_{r=1}^M F_r(k)}{ML^2},
\end{equation}
where $F_r(k)$ is the number of agents having $k$-th chunk of knowledge [i.e. with $c_k(x,y,t\to\infty)=1$] in $r$-th simulation.

In this experiment, we can look at the effectiveness of knowledge transfer more globally, i.e. in the whole system (the whole organization).
The more chunks of knowledge ($K$) is required in the organization, the greater must be the initial concentration of knowledge in the organization ($p$) to allow almost all agents acquire the required knowledge [see Fig.~\ref{F:fci_vs_Kp}].
Moreover, comparing the three models A, B and C---and basing on results presented in Fig.~\ref{F:fci_vs_Kp}---the transfer of knowledge is more effective for model B than A.
A lower values of initial concentration of chunks of knowledge $p$ are required to ensure the global saturation of skills in organization ($\langle f\rangle=1$) when comparing models B and A [cf. Fig.~\ref{F:fci_vs_Kp}(d) and Fig.~\ref{F:fci_vs_Kp}(b)].
For model B these values increse with the number of chunks of knowledge $K$ desired in the organization.
Obviously the best results, as in the previous experiments, are obtained for the situation C [cf. Fig.~\ref{F:fci_vs_Kp}(f) and Fig.~\ref{F:fci_vs_Kp}(d)].

\begin{figure}[htb]
\centering
\psfrag{p}{$p$}
\psfrag{K=}{$K=$}
\psfrag{pC}{$p^*$}
\includegraphics[width=.41\textwidth]{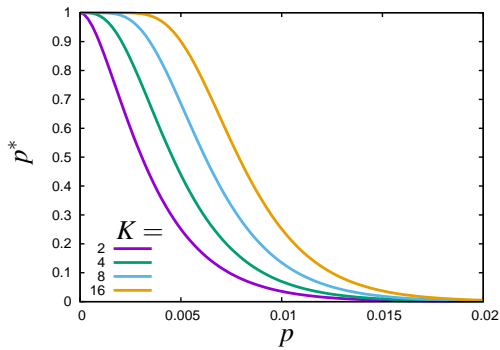}
\caption{\label{F:pC_vs_p}The dependencies $p^*(p)$ [Eq.~\eqref{eq:pC}] for $L=20$ and various values of $K$.}
\end{figure}

As it was mentioned in Ref.~\cite{Kowalska-2017}, for model A (Sec.~\ref{modelA}), ``the dependence $\langle f\rangle$ on $p$ increases to a certain threshold values of $p$ [\ldots] then it decreases, and it increases again [\ldots] We see [\ldots], that for $L=20$ and $K\ge 7$ a drop in $\langle f\rangle$ is absent.''
The surprising effect of decreasing the effectiveness of the knowledge transfer with increasing the initial concentration $p$ of chunks of knowledge among agents for $K<7$ [cf. Figs.~\ref{F:fci_vs_Kp}(a-b)] was directly associated with and explained by the assumed rules of the knowledge transfer described in Ref.~\cite{Kowalska-2017}.
In particular, for the minimal set ($K=2$) of required chunks of knowledge one may evaluate the mean-field-like probabilities of finding pairs of agents allowing for knowledge transfer or blocking it~\cite{Kowalska-2017}.
The latter vanishes for model B and for model C, what results in monotonically non-decreasing $\langle f\rangle$ vs. $p$ dependencies for model B and model C as presented in Figs.~\ref{F:fci_vs_Kp}(c-d) and \ref{F:fci_vs_Kp}(e-f). 
Moreover, when sending the chunks of knowledge is not restricted only to the ``smarter'' agents (model C), the quantitative difference on $\langle f\rangle$ vs. $p$ curves for various values of $K$ disappears.
As we can deduce from Fig.~\ref{F:fci_vs_Kp}(e-f)---when model C is assumed---the agents are able to acquire all chunks of knowledge desired by the organization independently on the level $K$ of the knowledge required in the organization if only initial concentration of chunks of knowledge is sufficiently large.
Moreover, in model C, agents will not acquire all of $K$ chunks of knowledge only when some chunk of knowledge will not be available initially in the organization, i.e. when
\begin{equation}
\label{eq:cond}
\exists k\,\forall x,y\colon c_k(x,y,0)=0 \text{ and } (x,y)\in\mathbb{N}\times\mathbb{N},\, 1\le x,y\le L.
\end{equation}
The condition~\eqref{eq:cond} is fulfilled with probability
\begin{equation}
\label{eq:pC}
p^*=1-\left((1-p)^{L^2}\right)^K,
\end{equation}
which is the probability that at least one column in random binary matrix with $K$ columns and $L^2$ rows will contain only zeros when matrix elements are ``1'' with probability $p$ and ``0'' with probability $(1-p)$.
The dependencies $p^*(p)$ for $L=20$ and various values of $K$ are presented in Fig.~\ref{F:pC_vs_p}.
As we can see, the probabilities that some chunks of knowledge will be initially missing in organization with $L^2=400$ members are marginal if $p>2\%$.

\section{\uppercase{Discussion and Conclusion}}

In this work we are looking for a better way to model knowledge transfer in an organization, i.e. we are searching for models giving better efficiency and better effectivity of knowledge transfer than those obtained and described in Ref.~\cite{Kowalska-2017}.
Therefore, we have investigated a CA model to answer the question if differences on the microscopic scale between the three versions of the knowledge transfer model manifest in differences in the efficiency and speed of knowledge transfer. 

Many studies show the importance of social interaction among organizational members in knowledge exchange~\cite{Chen-2007,Ibarra-1993,Tsai-2002}. 
It should be noted, that these interactions between organization members are mainly informal and these interactions are the main source of influence in organizations~\cite{Ibarra-1993}.
Therefore, the proposed models base on informal relations between the organization members.
Moreover, apart from the stock of knowledge, the efficient knowledge transfer requires the strong disseminative capacity of knowledge senders and the absorptive capacity of knowledge receivers~\cite{Tang-2011}.
The abilities of people \emph{i}) to efficiently and effectively communicate, \emph{ii}) to spread knowledge in a way that other people can understand information accurately, \emph{iii}) and to use knowledge in practice are very important. 
The lack of competence, skills and language efficiency of knowledge senders makes the transfer of their knowledge to others rather difficult~\cite{Cabrera-2003}.
It should be added, that in this knowledge transfer process, leaders also play an important role \cite{Girdauskiene-2012}.
They become informal leaders (distributed leadership), when leadership is understood as a common social process resulting from the interaction of multiple entities~\cite{Uhl-Bien-2006}.
In this perspective, the way of leaders interaction with others is more important than the nature of their leadership roles, responsibilities or functions~\cite{Harris-2012}.
In connection with this, we define the rule of knowledge transfer as a transfer of knowledge from the people who have more of knowledge chunks (leaders) to people with less of knowledge chunks (followers).

In this paper we have assumed three situations in which the sender is eligible to transfer knowledge to the recipient:
\begin{itemize}
\item when the sender has exactly one more chunks of knowledge than recipient (model A),
\item when the sender has more chunks of knowledge than recipient (model B),
\item when the sender has the same or greater number of chunks of knowledge than recipient (model C).
\end{itemize}

As we have shown, the differences between model A and B are almost negligible (are not large), although the model B gives a better result than model A.
In the model B, we assume that the transfer occurs when sender has more chunks of knowledge than recipient (i.e. when the sender generally has a greater knowledge of the recipient).
It seems to be better than the assumptions in the model A, because, it seems more natural that people with more knowledge (knowledge leaders) share it with others.
The knowledge transfer is most effective in the model C. 
In this situation, the senders are also people who have the same number of chunks knowledge as the recipient (i.e., they may be at a similar level of knowledge).

The simulations, for all three models, show a significant role of initial concentration of knowledge chunks in the transfer process.
Organizations should therefore carry out training courses for their employees in order to increase the initial knowledge in the organization. 
However, as it was shown earlier, the most important is the way to knowledge transfer.
The best results are obtained  when the transfer of knowledge is possible when the recipient have the same or lower level of knowledge from the broadcasters (model C), because the knowledge transfer occurs more frequently.
Therefore, in case of spontaneous knowledge transfer, managers should develop and support informal contacts between employees to reduce the social distance between them.

\section*{\uppercase{Acknowledgments}}

This research was supported by \href{https://www.ncn.gov.pl/?language=en}{National Science Center} (NCN) in Poland (grant no. UMO-2014/15/B/HS4/04433) and partially by Polish Ministry of Science and Higher Education.

\vfill
\bibliographystyle{apalike}
{\small
\bibliography{opus8}}

\end{document}